\documentclass[
reprint,
nofootinbib,
amsmath,
amssymb, 
aps,
pra,
]{revtex4-1}
\usepackage{dcolumn}
\usepackage{bm}
\usepackage{color}
\usepackage[T1]{fontenc}
\usepackage{mathrsfs,amsfonts,dsfont}
\usepackage{amstext}
\usepackage{mathtools}
\usepackage{enumitem}
\usepackage{textcomp}
\newcommand{\tr}[1]{\text{tr}\left(#1\right)}
\newcommand{\ket}[1]{\left\vert#1\right\rangle}
\newcommand{\bra}[1]{\left\langle#1\right\vert}
\newcommand{\ghz}{\text{GHZ}_N}

\begin{document}
\preprint{APS/123-QED}
\title{
Comment on "Fully device-independent conference key agreement" \\
$[$Phys. Rev. A \textbf{97}, 022307 (2018)$]$
}
\author{Timo Holz}\email{holzt@uni-duesseldorf.de}
\author{Daniel Miller}
\author{Hermann Kampermann}
\author{Dagmar Bru\ss}
\affiliation{Institut f\"ur Theoretische Physik $\MakeUppercase{\romannumeral3}$, Heinrich-Heine-Universit\"at D\"usseldorf, D-40225 D\"usseldorf, Germany}
\date{\today}
\begin{abstract}
In this manuscript we discuss the device-independent conference key agreement (DICKA) protocol [Phys. Rev. A 97, 022307 (2018)]. We show that the 
suggested honest implementation fails, because perfect correlated measurement results and the required Bell-inequality violation cannot be achieved simultaneously, 
in contradiction to what is claimed. We further show via semidefinite programming that there cannot exist {\em any} suitable honest implementation in the tripartite 
setting, rendering the DICKA protocol incomplete.
\end{abstract}
\maketitle
In Ref.~\cite{DICKA}, Ribeiro \emph{et al.} proposed a protocol to generate a secret key among multiple parties, 
called device-independent conference key agreement (DICKA). The security proof crucially depends on the observation of genuine multipartite entanglement 
certified by a particular violation of a multipartite Bell inequality, the Mermin-Ardehali-Belinski{\u\i}-Klyshko (MABK) inequality~\cite{mermin, ardehali, belinskii}. 
Here, we analytically prove that the honest implementation of the DICKA protocol cannot yield a non-zero secret-key rate for an odd number of parties and 
provide numerical evidence that the first non-trivial even-numbered case fails as well. Finally, we use semidefinite programming (SDP) to prove that there cannot 
exist {\em any} honest implementation that leads to a non-vanishing secret-key rate for three parties, thus proving the incompleteness of the DICKA protocol. 
We use the same notation as Ref.~\cite{DICKA}.
\paragraph*{MABK inequality:}
Consider a Bell setup with $N$ parties called Paul$_i$ with two dichotomic observables $P_0^i, P_1^i$ $\forall i\in[N]\coloneqq\{1,\dots,N\}$. We 
require an explicit expression for the odd-partite MABK operator. Let $\mathbb{F}_2=\{0,1\}$ denote the finite field with two elements, from which we obtain 
the vector space $\mathbb{F}_2^N$ of bit strings of length $N$. We define the Hamming weight
\begin{align}
H(\bm{x}) \coloneqq \big\vert \{ 1\le i \le N\,\, \vert\,\, x_i =1 \} \big\vert 
\end{align}
of a bit string $\bm{x} = (x_1,\dots,x_N)$. For now, let $N$ be an odd integer and define the set 
\begin{align}
\mathcal{L}_{N} \coloneqq \left\{\bm{x}\in\mathbb{F}_2^N \,\, \bigg\vert\,\, H(\bm{x}) = \frac{N-1}{2}\mod{2} \right\}, \label{setX}
\end{align}
i.e., if $(N-1)\slash2$ is odd (even) the set $\mathcal{L}_{N}$ contains all bit strings $\bm{x}$ with an odd (even) number of bits $1$. 
\begin{flushleft}
\textbf{Proposition} -- Let $N\ge3$ be odd. An explicit form of the $N$-MABK operator is given by
\begin{align}
 MK_N = \frac{1}{\mathcal{N}_N}\sum\limits_{\bm{x}\in\mathcal{L}_{N}} (-1)^{\xi_N(\bm{x})} \bigotimes\limits_{i=1}^NP_{x_i}^{i}, \label{mabk_nonrecursive}
\end{align}
where $\xi_N(\bm{x})\coloneqq \frac{N-1}{4}-\frac{H(\bm{x})}{2}$ and $\mathcal{N}_N\coloneqq2^{\frac{N-1}{2}}$.
\end{flushleft}
A single application of the recursion rule in Eq.~($8$) of Ref.~\cite{DICKA}, yields the MABK operator for $N$ even. 
For all $N\ge3$ the $N$-MABK inequality is given by the corresponding MABK operator $MK_N$, according to:
\begin{align}
 \mathcal{MK}_N \coloneqq \left\vert\tr{MK_N \rho_{\mathcal{P}_{(1\dots N)}}}\right\vert \le 2^{\frac{m-1}{2}},  \label{mabkinequality}
\end{align}
where $\rho_{\mathcal{P}_{(1\dots N)}}$ denotes the quantum state shared among all $N$ parties and $m\in[N]$ indicates the maximum number of parties that are 
entangled via $\rho_{\mathcal{P}_{(1\dots N)}}$. A violation of the bound for $m = N-1$ certifies genuine $N$-partite entanglement~\cite{wernerBellPPT}, which is 
crucial for the security proof of the DICKA protocol.\\
There are $\vert \mathcal{L}_N \vert = 2^{N-1}$ different operators in the sum of Eq.~\eqref{mabk_nonrecursive}. Thus, the $N$-MABK value 
$\mathcal{MK}_N$  contains $E_N=2^{N-1}$ different expectation values. For general $N\ge2$, the number of different expectation 
values $E_N$ and the normalization factor $\mathcal{N}_N$ are given by 
\begin{align}
E_N = 2^{2\lfloor \frac{N}{2}\rfloor}, \,\,\text{and}\,\,\, \mathcal{N}_N = 2^{\lfloor\frac{N}{2}\rfloor}. \label{properties}
\end{align}
\paragraph*{DICKA protocol and honest implementation:}
Alice has a measurement device with two inputs $X\in\{0,1\}$, and each Bob$_k$ has three inputs $Y_{(k)} \in\{0,1,2\}$ for 
$k\in[N-1]$. The DICKA protocol consists of two different types of measurement rounds, one for key generation (type $0$), where 
$(X,Y_{(1\dots N-1)})=(0,2,\dots,2)$, and one for parameter estimation (type $1$), where $X,Y_{(k)}$ are chosen uniformly at random from $\{0,1\}$. In the honest 
implementation and in the asymptotic limit, the parties have access to infinitely many copies of the pure $N$-Greenberger-Horne-Zeilinger (GHZ) state 
$\ghz\coloneqq\ket{\text{GHZ}_N}\bra{\text{GHZ}_N}$, with $\ket{\text{GHZ}_N}\coloneqq\frac{\ket{0}^{\otimes N}+\ket{1}^{\otimes N}}{\sqrt{2}}$, 
which are distributed to the parties. Alice and the Bobs measure the observables (see section between Protocol~$2$ and Thm.~$4$ of the Ref.~\cite{DICKA}): 
\begin{enumerate}[label=(\roman*)]
 \item For $X=0$ ($X=1$) Alice's observable is $\sigma_z$ $(\sigma_x)$.
 \item For type-$0$ measurement rounds, i.e., for $Y_{(k)} = 2$ all Bobs measure the observable $\sigma_z$. And for type $1$, i.e., $Y_{(k)}\in\{0,1\}$ 
 they measure observables "that are defined by a strategy that maximally violates the $N$-MABK inequality when the measurements are performed on an $N$-GHZ state."
\end{enumerate}
We want to emphasize the following remarks. 
First, note that in any DI quantum key distribution (QKD) protocol, at least one party, say Alice, is obliged to incorporate at least one measurement setting that is 
used for key generation rounds also in the parameter estimation rounds, to detect a potential pre-programming of the devices by the adversary. 
And second, in order to minimize the error correction information that is publicly communicated, and given that the $N$-GHZ state is measured, 
every party necessarily needs to measure the observable $\sigma_z$ in type-$0$ rounds of the protocol, see Thm.~$1$ of Ref.~\cite{NQKD}. 
Therefore, Alice has to use $A_0=\sigma_z$ in both types of measurements. We claim that under these conditions there exist no measurement settings for the Bobs such 
that the $N$-MABK value exceeds the bound $2^{\frac{m-1}{2}}$ for $m=N-1$ in Ineq.~\eqref{mabkinequality}, at least for odd $N$. Hence, 
the security of the DICKA protocol cannot be guaranteed.\\
Let $\mathcal{P}_N$ denote the $N$-qubit Pauli group, we define
\begin{align}
\mathcal{S} \coloneqq \Big\{S \in \mathcal{P}_N \,\,\Big\vert\,\, S \ket{\text{GHZ}_N} = \ket{\text{GHZ}_N} \Big\}, 
\end{align}
i.e., $\mathcal{S}$ denotes the stabilizer group of the $N$-GHZ state. The group $\mathcal{S}$ is generated by the $N$ independent 
operators
\begin{subequations}
\label{generators}
\begin{align}
G_1&\coloneqq \sigma_x^{\otimes N}, \quad \text{and for all } j \in \{2,\dots,N\}: \\ 
G_j&\coloneqq \bigotimes\limits_{i=1}^{j-2}\mathds{1}_2^{(i)}\otimes \sigma_z^{(j-1)}\otimes\sigma_z^{(j)}\otimes\bigotimes\limits_{i=j+1}^{N}\mathds{1}_2^{(i)}
\end{align}
\end{subequations}
where the superscript denotes the corresponding subsystems. In general, the projector of any stabilizer state can be written as the normalized sum 
of all of its stabilizer operators~\cite{stabilizer0,stabilizer00}. We obtain for $\ghz$ and with $\bm{s}\coloneqq(s_1,\dots,s_N)$ the representation:
\begin{align}
 \ghz   =  \frac{1}{2^N}\sum\limits_{\bm{s}\in\mathbb{F}_2^N}\left(\sigma_x^{s_1}\right)^{\otimes N} &\big(\sigma_z^{s_2} \otimes  \nonumber
\sigma_z^{s_2+s_3} \otimes    
\dots \\ 
&\dots\otimes \sigma_z^{s_{N-1}+s_N} \otimes \sigma_z^{s_N}
\big). \label{generalGHZ}
\end{align}
\paragraph*{No-go theorem for $N$ odd:}
With the general form of the pure $N$-GHZ state in Eq.~\eqref{generalGHZ} and the properties of the $N$-MABK inequality, cf. Eq.~\eqref{properties}, 
we state our initial claim:
\begin{flushleft}
\textbf{Theorem $\bm{1}$} -- Let $N\ge3$ be odd and let the $N$ parties perform the honest implementation of the DICKA protocol. 
Then, the $N$-MABK value cannot exceed the bound that certifies genuine multipartite entanglement among all $N$ parties.
\end{flushleft}
\emph{Proof} -- 
Let $N\in\mathbb{N}$ be an odd integer and let 
\begin{align}
B_{Y_{(k)}}^{(k)} \coloneqq \bm{\beta}_{Y_{(k)}}^{(k)\,T} \bm{\sigma}, \,\,\,\, \forall k \in [N-1], \,\, Y_{(k)}\in\{0,1\} \label{bobobservable}
\end{align}
be a general qubit measurement, where 
\begin{subequations}
\label{generalQubit}
\begin{align}
\bm{\beta}_{Y_{(k)}}^{(k)} &\coloneqq \left(\beta_1,\beta_2, \beta_3\right)_{Y_{(k)}}^{(k)\,T}\equiv \left(\beta_x,\beta_y,\beta_z\right)_{Y_{(k)}}^{(k)\,T}, \\
\bm{\sigma} &\coloneqq \left(\sigma_1,\sigma_2,\sigma_3\right)^T\equiv \left(\sigma_x,\sigma_y,\sigma_z\right)^T
\end{align}
\end{subequations}
denote normalized Bloch vectors and a vector that contains the Pauli matrices. 
Recall that the product of Pauli matrices is given by:
\begin{align}
 \sigma_j \sigma_k = \delta_{j,k}\mathds{1}_2 + i \sum\limits_{l=1}^{3}\epsilon_{jkl}\sigma_l , \label{paulirelation}
\end{align}
where $\delta_{j,k}$ and $\epsilon_{jkl}$ denote the Kronecker delta and the Levi-Civita tensor, respectively. 
With Eq.~\eqref{generalGHZ} and for $A_0 = \sigma_z$, we obtain for the expectation value:
\begin{align}
\nonumber 
\left\langle\! A_0 \bigotimes\limits_{k=1}^{N-1} B_{Y_{(k)}}^{(k)}\!\right\rangle
=\!\!\!\sum\limits_{\bm{s}\in\mathbb{F}_2^N}\!\frac{2}{2^N}\delta_{s_1,0}&\,\delta_{s_2,1} 
\text{tr}\Big(B_{Y_{(1)}}^{(1)}\!\sigma_x^{s_1}\!\sigma_z^{s_2+s_3} \otimes\\ 
&\dots\otimes\! B_{Y_{(N\!-\!1)}}^{(N\!-\!1)}\!\sigma_x^{s_1}\!\sigma_z^{s_N}\!\Big)
\label{expectationvalue}
\end{align}
for all $Y_{(k)}\in\{0,1\}$, where we used Eq.~\eqref{paulirelation} and the fact that Pauli matrices are traceless to establish
\begin{align}
\tr{\sigma_z\sigma_x^{s_1}\sigma_z^{s_2}}=2\delta_{s_1,0}\delta_{s_2,1}. \label{Alice}
\end{align}
Therefore, only operators with components $s_1=0$ and $s_2=1$ in Eq.~\eqref{generalGHZ}, yield a non-vanishing contribution to the expectation 
value in Eq.~\eqref{expectationvalue}. Now consider 
\begin{align}
 B_{Y_{(1)}}^{(1)}\sigma_x^{s_1}\sigma_z^{s_2+s_3}=\sum\limits_{i=1}^3 \beta_{i, Y_{(1)}}^{(1)}\sigma_i\sigma_x^{s_1}\sigma_z^{s_2+s_3}
\end{align}
and note that we only get a non-vanishing contribution to the expectation value in Eq.~\eqref{expectationvalue}, if there remains no non-trivial Pauli matrix in this 
expression. As $s_1=0$ and $s_2=1$, we see that $s_3=0$ needs to hold. 
Repeating this argument reveals, that the only term in Eq.~\eqref{generalGHZ} that potentially gives a non-zero contribution 
to the expectation value, is the bit string $\bm{s}$ with alternating entries of $0$ and $1$. Here, we observe a 
fundamental difference between odd and even numbers $N$. For $N$ odd, the alternating pattern in $\bm{s}$ implies $s_N=0$, thus for the 
observable of Bob$_{N-1}$ in Eq.~\eqref{expectationvalue}, we obtain $B_{Y_{(N-1)}}^{(N-1)}\sigma_x^{s_1}\sigma_z^{s_N}= B_{Y_{(N-1)}}^{(N-1)}\mathds{1}$, 
which is traceless. Hence, the expectation value in 
Eq.~\eqref{expectationvalue} necessarily vanishes if $A_0=\sigma_z$, i.e., for all $N\ge3$ odd, we obtain:
\begin{align}
 \left\langle \sigma_z \otimes \bigotimes\limits_{k=1}^{N-1} B_{Y_{(k)}}^{(k)}\right\rangle_{\ghz} = 0. \label{exp_nodd}
\end{align}
The structure of the $N$-MABK Ineq.~\eqref{mabkinequality} is such that half of the expectation values $E_N$ include the observable $A_0=\sigma_z$. Thus, 
only $E_N/2$ non-zero expectation values, each of them upper bounded by $+1$, can be present. A multiple application of the 
triangle inequality leads to
\begin{align}
\mathcal{MK}_N = \left\vert\tr{MK_N \rho}\right\vert \le \frac{1}{2}\frac{E_N}{\mathcal{N}_N} = 2^{\frac{N-3}{2}} \label{generous}
\end{align}
as a generous upper bound on the $N$-MABK value. A comparison with the bound that certifies genuine $N$-partite entanglement, 
cf. Ineq.~\eqref{mabkinequality} for $m=N-1$, reveals 
\begin{align}
\mathcal{MK}_N \le 2^{\frac{N-3}{2}} < 2^{\frac{N-2}{2}},
\end{align}
which finishes the proof. \hfill $\square$
\paragraph*{The even-partite case:}
The even-numbered analogon to Eq.~\eqref{exp_nodd} is given by:
\begin{align}
\left\langle \sigma_z \otimes \bigotimes\limits_{k=1}^{N-1} B_{Y_{(k)}}^{(k)}\right\rangle_{\ghz} \!\!= 
 \prod\limits_{k=1}^{N-1}\beta_{z,Y_{(k)}}^{(k)},  \label{exp_n4}
\end{align}
which is in general non-vanishing and thus prohibits an analogous analytical proof of a similar no-go theorem for even $N$. 
Numerical optimization procedures however, can be utilized to find the maximum possible $N$-MABK value, where the maximization is done 
over all Bloch components of all observables but $A_0$ under the constraints of normalization for each Bloch vector. 
For $N=4$, the maximization yields an upper bound of $1$ for the maximum $4$-MABK value, which constitutes the classical bound. Thus, we obtain 
numerical evidence, that in the first non-trivial even-numbered case of multipartite DICKA, the honest implementation fails as well.
\paragraph*{A device-independent generalization:}
The results presented so far hint at fundamental problems of the DICKA protocol employing the MABK inequality. 
More precisely, perfect classical correlations and certified genuine multipartite entanglement via an MABK-inequality violation seem to be incompatible. 
So let us move away from the honest implementation. All we demand is that there exists a set of observables and a quantum state 
that perfectly correlate the measurement results of all parties in type-$0$ measurement rounds of the DICKA protocol. 
Besides this premise, no specific structure on the quantum state and the (projective, dichotomic) measurements are imposed. 
In this sense, it is a DI way to reinforce the results presented so far. To do this, we employ the Navasqu\'{e}s-Pironio-Ac\'in (NPA) 
hierarchy~\cite{npa}, whose generality comes at the cost of being numerically expensive. We thus restrict the discussion to $N=3$ parties. Note, however that the 
extension to larger $N$ is straightforward. 
\begin{flushleft}
\textbf{Theorem $\bm{2}$} -- Given a $3$-partite quantum state $\rho$ and a set of observables $\left(A_X,B_{Y_{(1)}}^{(1)},B_{Y_{(2)}}^{(2)}\right)$ 
that lead to perfectly correlated measurement results among all parties in type-$0$ measurement rounds of the DICKA protocol, then
\begin{align}
\mathcal{MK}_3 \le 2^{\frac{1}{2}}
\end{align}
holds, i.e., the $3$-MABK value cannot exceed the bound that certifies genuine multipartite entanglement.
\end{flushleft}
To show this theorem, let without loss of generality the
indices $(X,Y_{(1)},Y_{(2)}) = (0,2,2)$ indicate the set of inputs that yields perfect classical correlations. Let $A_0^\pm$, $B_{2}^{(1)\,\pm}$, and 
$B_{2}^{(2)\,\pm}$ denote the projectors onto the $\pm1$ eigenstate of the corresponding observables 
and define:
\begin{align}
\mathcal{C}
\coloneqq A_0^+\otimes B_{2}^{(1)\,+}\!\!\otimes B_{2}^{(2)\,+} \!+ A_0^-\otimes B_{2}^{(1)\,-}\!\!\otimes B_{2}^{(2)\,-}.
\end{align}
The solution of the following SDP, for which we use Ref.~\cite{wittek}, is the maximum $3$-MABK value in this general setting, 
subject to the constraint of perfect correlations.
\begin{alignat}{1} \label{SDP}
\max\limits_{A_X, B_{Y_{(1)}}^{(1)}, B_{Y_{(2)}}^{(2)},\rho} \,\,\,	& \big\vert \tr{MK_3\rho} \big\vert \\  
\text{subject to: }	&\,\,\,\tr{\mathcal{C}\rho}=1 \nonumber
\end{alignat}
The upper bound on the $3$-MABK value obtained via the solution of the SDP~\eqref{SDP}
coincides with the bound that certifies genuine multipartite entanglement within numerical precision. Thus, there cannot exist a quantum state and a set of 
observables that simultaneously perfectly correlate all parties and lead to the required $3$-MABK-inequality violation, which proves Thm.~$2$. The $N$-partite 
generalization of the SDP~\eqref{SDP} can be carried out for arbitrary integers $N$ at a proper hierarchy level.
\paragraph*{Conclusion:}
We presented an analytical proof that the honest implementation of the DICKA protocol proposed in Ref.~\cite{DICKA} fails for an odd number $N$ of parties and provided 
numerical evidence that the protocol fails in the first non-trivial even-numbered case as well. We furthermore proved via SDP that there cannot exist {\em any} honest implementation 
of the DICKA protocol relying on the violation of the MABK inequality for $N=3$ parties, thus proving its incompleteness. We finally conjecture that the $N$-partite 
generalization of Thm.~$2$ holds also true, which suggests that there cannot exist {\em any} honest implementation of the $N$-partite DICKA protocol that leads to 
a non-zero secret-key rate. 
\begin{flushleft}
The authors acknowledge support from the Federal Ministry of Education and Research (BMBF, Project Q.Link.X).
\end{flushleft}
\bibliography{CommentDICKA_TimoHolz}
\end{document}